\documentclass[useAMS,usenatbib]{mn2e}

\usepackage{dcolumn}% Align table columns on decimal point
\usepackage{graphics,epsfig,amssymb}
\usepackage{times}

\newcommand{\kms}{km~s$^{-1}$}
\newcommand{\cm}{cm$^{-2}$}

\newcommand{\NHI}{N_{\rm HI}}

\newcommand{\ts}{{\rm T_s}}
\newcommand{\pks}{PKS~B0438$-$436}
\title[HI 21cm Absorption at $z \sim 2.347$ towards PKS~B0438$-$436]{HI 21cm absorption at $z \sim 2.347$ towards PKS~B0438$-$436}
\author[Kanekar et al.]{N.~Kanekar$^1$\thanks{E-mail: nkanekar@aoc.nrao.edu (NK); 
rsubrahm@rri.res.in (RS); sarae@uvic.ca (SLE); wendy.peters@nrl.navy.mil (WML); 
chengalu@ncra.tifr.res.in (JNC)}, R.~Subrahmanyan$^2$, S.~L.~Ellison$^3$, W.~M.~Lane$^4$, 
J.~N.~Chengalur$^{2,5}$ \\
$^{1}${}National Radio Astronomy Observatory, Socorro, NM 87801, USA; 
$^{2}${}Australia Telescope National Facility, Epping, Australia; 
$^{3}${}University of Victoria, \\ BC, Canada; 
$^{4}${}US Naval Research Laboratory, Washington, USA; 
$^{5}${}National Centre for Radio Astrophysics, Pune, India 
}

\begin{document}

\date{Received mmddyy/ accepted mmddyy}

\maketitle

\label{firstpage}

\begin{abstract}
We report the detection of redshifted HI~21cm absorption in
the $z \sim 2.347$ damped Lyman-$\alpha$ absorber (DLA) towards \pks,
with the Green Bank Telescope. This is the second-highest redshift at 
which 21cm absorption has been detected in a DLA. The absorption extends over 
$\sim 60$~\kms~and has two distinct components, at $z = 2.347477 (12)$ 
and $z = 2.347869 (20)$. A similar velocity structure is seen in optical 
metal lines, although the peak absorption here is offset by $\sim 11$~\kms~from 
the peak in the 21cm line. We obtain a high spin temperature $\ts \sim (886 \pm 248) 
\times (f/0.58)$~K, using a covering factor estimated from 2.3~GHz VLBI data.
However, the current data cannot rule out a low spin temperature.
The non-detection of CO~3--2 absorption places the upper limit 
N$_{\rm CO} < 3.8 \times 10^{15} \times (T_x/10)$~\cm~on the CO column density.
\end{abstract}

\begin{keywords}
galaxies: high redshift -- galaxies: ISM -- radio lines: galaxies
\end{keywords}

\section{Introduction}
\label{intro}

The highest HI column density absorption systems in quasar spectra, the 
damped Lyman-$\alpha$ absorbers (DLAs), are important in the context 
of galaxy evolution as they are believed to be the precursors of present-day 
galaxies \citep{wolfe86}. Understanding the nature of a typical DLA and its 
evolution with redshift is perhaps the key to understanding normal galaxy 
evolution. However, despite a tremendous observational effort over the last two 
decades, physical conditions in high $z$ DLAs are still the subjects of much debate
(see \citet{wolfe05} for a recent review).

HI 21cm absorption studies of DLAs towards radio-loud background sources provide 
crucial information on the nature of the absorbing galaxy (e.g. \citealt{kanekar04}). 
21cm lines are usually optically thin and trace the kinematics of the neutral gas;
they can thus be used to obtain the velocity field of the absorber and its 
HI and dynamical masses \citep{briggs01}. In some cases, the gas kinetic 
temperature can be directly measured with high resolution 21cm spectroscopy 
(e.g. \citealt{lane00}). Comparisons between the redshifts of 
21cm and metal-line or molecular absorption allow estimates of
changes in fundamental constants (e.g. \citealt{tzanavaris05}). 
Finally, the 21cm equivalent width can be combined with the HI column density 
to obtain the spin temperature $\ts$ of the absorbing gas (e.g. \citealt{kanekar03}); 
this contains information on the distribution of the HI between phases at 
different temperatures, one of the issues of controversy referred to above 
\citep{kanekar03,wolfe03}.

A problem with using 21cm absorption to probe physical conditions in high $z$ DLAs
is that there are, at present, only three absorbers in the literature with confirmed 
detections of absorption at $z > 1$ \citep{wolfe79,wolfe81,wolfe85}, with one further 
tentative detection at $z \sim 3.38$ \citep{briggs97}. Despite 
deep searches, the vast majority of systems only yield upper limits on $\ts$ 
(e.g. \citealt{kanekar03}). This is very different from the situation at low redshifts, 
$z \lesssim 0.8$, where 14 intervening 
21cm absorbers are known (e.g. \citealt{brown73,kanekar01a,lane02}). 
We have therefore been carrying out deep searches for 21cm absorption 
in a large sample of high redshift DLAs with the Green Bank Telescope (GBT)
and report here the detection of a new 21cm absorption system in the 
$z \sim 2.347$ DLA towards the $z \sim 2.863$ QSO, PKS~0438$-$436 
\citep{ellison01}.

\section{Observations and Data Analysis}
\label{obs}

\begin{figure*}
\centering
\epsfig{file=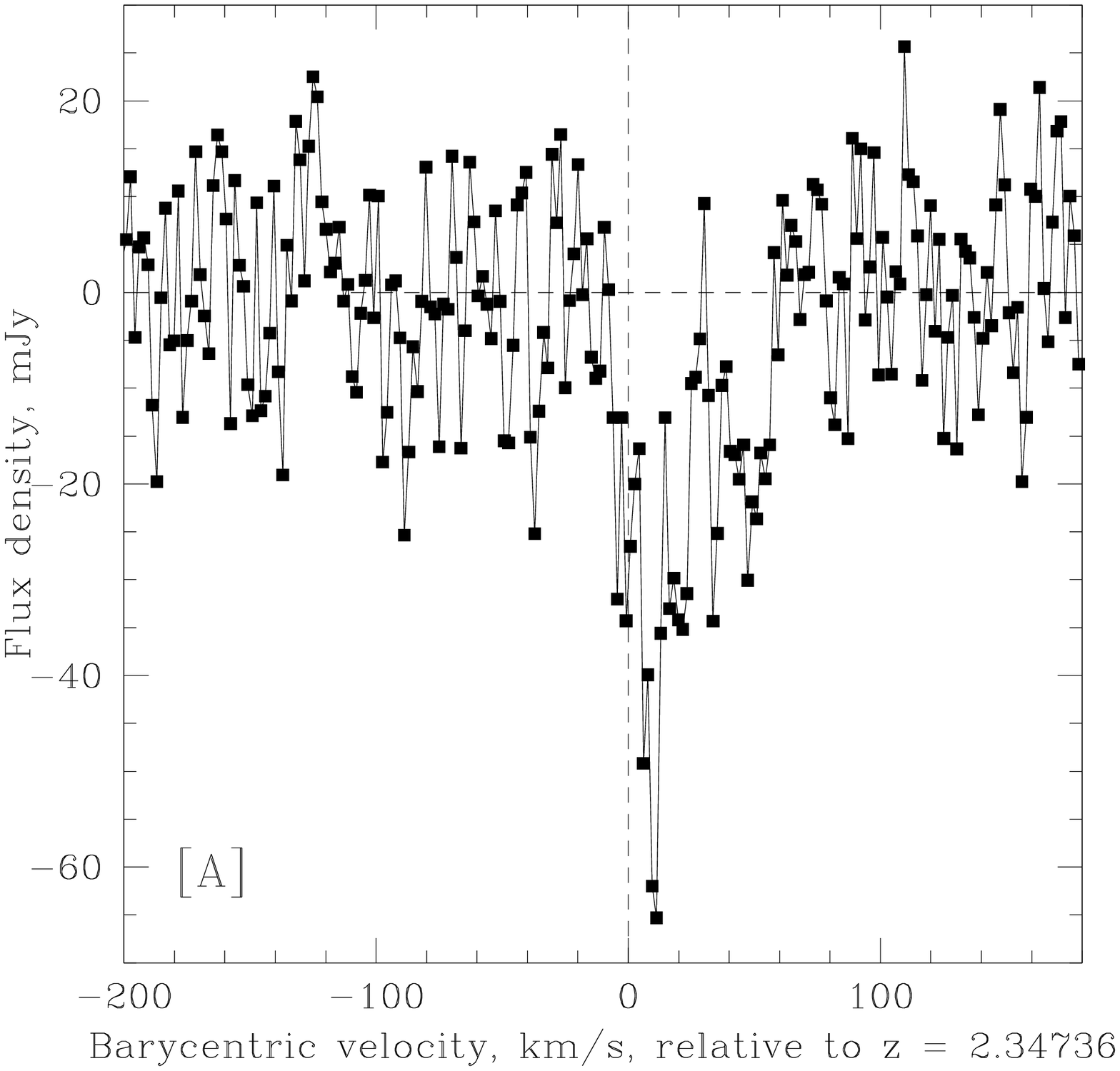,height=3.3truein,width=3.3truein}
\epsfig{file=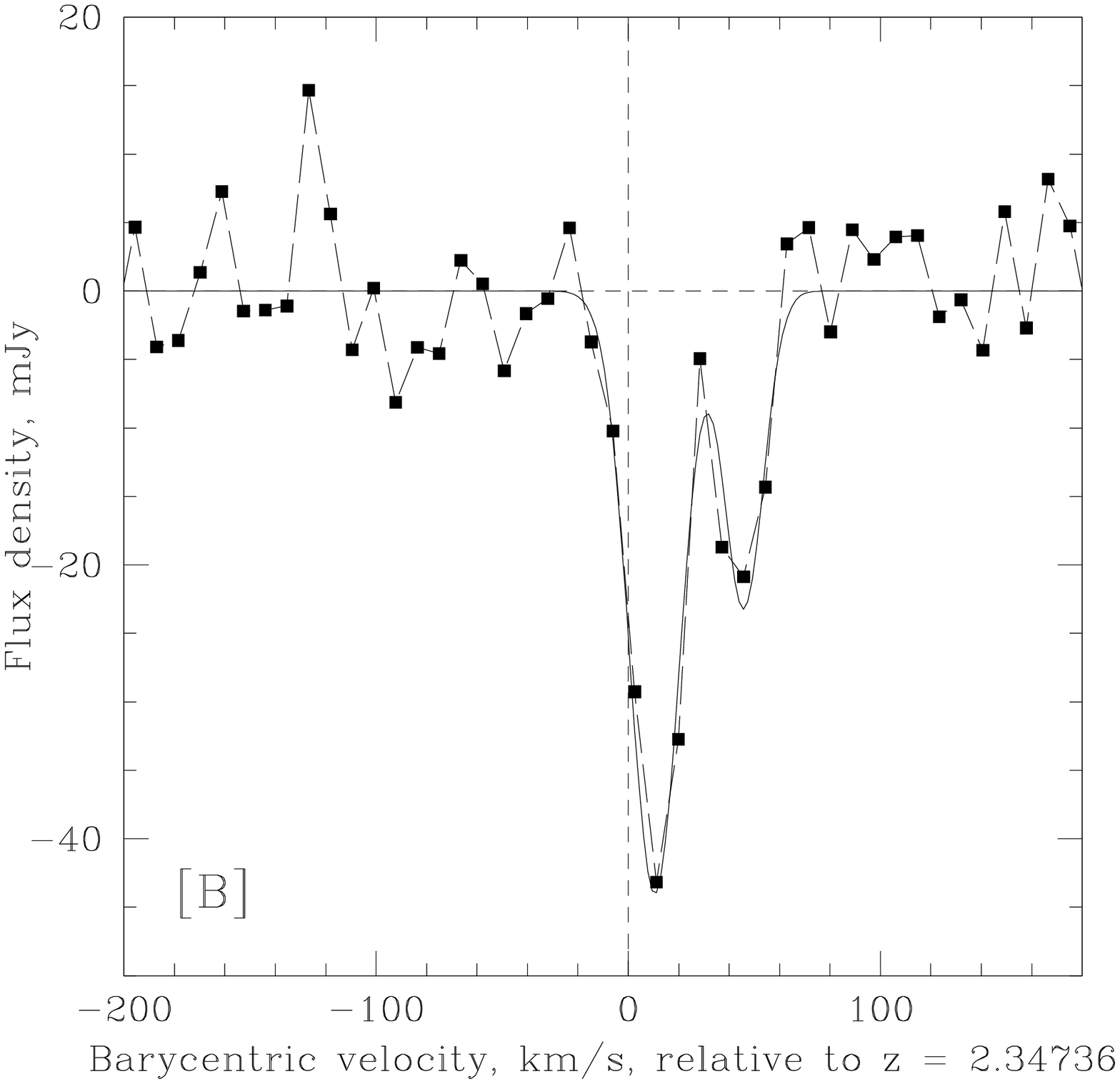,height=3.3truein,width=3.3truein}
\vskip -0.1in
\caption{Final GBT HI 21cm spectrum towards \pks~from August~2005, with flux
density (in mJy) plotted against barycentric velocity (in km/s), relative 
to $z = 2.34736$. [A]:~The spectrum at the original resolution of 
1.73~\kms. [B]:~The spectrum (dashed line) and 2-gaussian fit (solid line) 
after smoothing by 5~channels (and resampling) to a resolution of 
$\sim 8.6$~\kms. The QSO continuum flux density has been subtracted from both 
spectra.} 
\label{fig:hi}
\end{figure*}

The initial GBT observations of \pks~were carried out in January~2004, 
with the PF1-450~MHz receiver. These used the GBT Spectral Processor as 
the backend, with two linear polarizations and a 5~MHz bandwidth centred 
at 424.5~MHz and sub-divided into 1024 channels (giving a velocity 
resolution of $\sim 6.9$~\kms, after Hanning smoothing). A standard 
``position-switched'' mode was used, with ten-minute On-Off cycles 
made up of ten-second integrations. System temperatures were measured 
during the observations using a noise diode. The total on-source time 
was 30~minutes.

These observations resulted in the tentative detection of an absorption  
feature at $\sim 424.32$~MHz, close to the expected redshifted line frequency.
We hence re-observed \pks~in August~2005 to confirm the absorption,
at higher spectral resolution.  The observing and data recording procedures 
were the same as those of January~2004, except that a bandwidth of 1.25~MHz, 
centred at 424.3~MHz was used, sub-divided into 1024 channels (i.e.
a spectral resolution of $\sim 1.73$~\kms, after Hanning smoothing). The 
total on-source time was 80~minutes.

The data were analysed in the AIPS++ single dish package DISH, using standard 
procedures. No radio frequency interference (RFI) was found at or near the 
expected line frequency. After initial data-editing, the spectra were 
calibrated and averaged together, to measure the flux density of the background 
quasar, using a telescope gain of 2~K/Jy. This gave flux densities of 
$7.5 \pm 0.7$~Jy in January~2004 and $7.1 \pm 0.7$~Jy in August~2005, at 424~MHz, 
where the errors include those from confusing sources in the primary beam; 
(for comparison, \citet{large81} obtain $8.12\pm0.25$~Jy at 408~MHz). 
A second-order spectral baseline was then fit to each 10-second spectrum 
(during the process of calibration) and subtracted out; the frequency range 
covered in the fit was $\sim 423.7 - 424.7$~MHz, for both data sets. The 
residual 10-second spectra were then averaged together, to produce a spectrum 
for each epoch. The August~2005 spectrum was found to not be entirely flat 
at the end of this procedure; a second order polynomial was hence fit 
to this spectrum, excluding line channels, and subtracted out to produce 
the final spectrum. This has a higher sensitivity, a finer resolution and 
a better spectral baseline than the spectrum of January~2004, and will therefore
be used as the final 21cm spectrum towards \pks~in the discussion below. 

We also carried out a search for redshifted CO~3--2 absorption from the 
$z \sim 2.347$ DLA with the 3-mm receivers of the Australia Telescope 
Compact Array (ATCA), in October~2005. The antennas were in the EW214 
configuration, with a longest baseline of 214~m. A bandwidth of 64~MHz 
was used for the observations, centred at 103.315~GHz and divided into 
128~channels; this gave dual linear polarization spectra with an effective 
velocity resolution of 1.77~\kms. The observations consisted of 10-minute scans 
on \pks, interleaved with 2-minute scans on the bright nearby source 
PKS~B0454$-$463. The telescope pointing and system temperature calibrations 
were updated every 
hour; we obtained above-atmosphere T$_{\rm sys}$ in the range $300 - 350$~K in 
the different antennas, at the observing frequency. The planet Mars was 
observed at the start of the run to calibrate the absolute flux density 
scale. The total on-source time was 210~minutes.

The ATCA data were analysed in the software package {\sc MIRIAD}, using
standard procedures.  The complex gains and bandpass shapes of the ATCA 
antennas were derived from the visibility data on PKS~B0454$-$463 while 
the flux density scale was bootstrapped using the visibility amplitudes 
on Mars measured on the shortest (30-m) baseline (Mars is resolved 
on the longer ATCA baselines). The channel velocities towards \pks~were 
then shifted to the barycentric frame and the calibrated visibilities 
averaged to derive the final spectra.  The source flux density was measured 
to be 0.36~Jy, at a frequency of 103~GHz.

\section{Spectra and Results}
\label{sec:results}

\begin{figure}
\centering
\epsfig{file=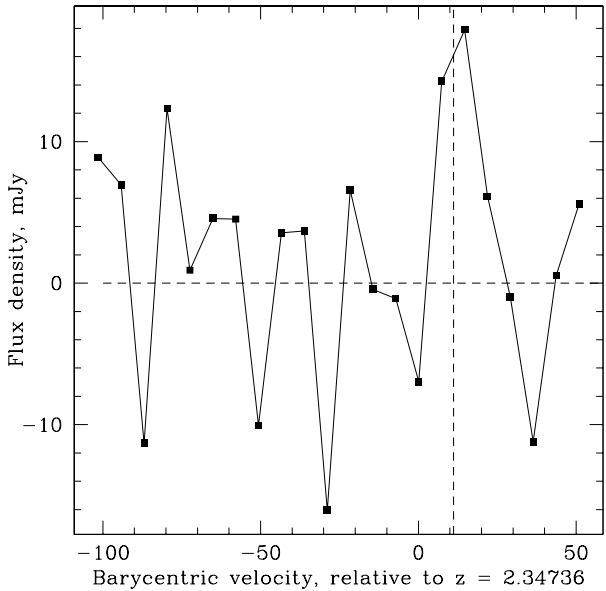,height=3.3truein,width=3.3truein}
\vskip -0.1in
\caption{ATCA 13~\kms~resolution CO~3--2 spectrum towards \pks, with flux density 
(in mJy) plotted against barycentric velocity (in \kms), relative to $z = 2.34736$. 
The QSO flux density has been subtracted from the spectrum. The dashed vertical 
line shows the CO~3--2 frequency corresponding to the peak 21cm line redshift. }
\label{fig:co}
\end{figure}

The final redshifted HI~21cm spectrum towards \pks~from August~2005 is shown 
in Figs.~\ref{fig:hi}[A] and [B], with flux density (in mJy) plotted against 
barycentric velocity (in km/s), relative to $z = 2.34736$, the redshift of the strongest
metal line absorption \citep{akerman05}. We adopt this as the systemic redshift 
of the DLA, indicated by the dashed vertical lines in the two panels of the
figure. Fig.~\ref{fig:hi}[A] shows the HI spectrum at the original 
velocity resolution of $\sim 1.73$~\kms, with an RMS noise of $\sim 10.4$~mJy per 
channel, while Fig.~\ref{fig:hi}[B] shows the spectrum smoothed by 
5~channels (and resampled), at a final resolution of $\sim 8.6$~\kms,
with an RMS noise of $\sim 4.7$~mJy per resampled channel.

The stronger 21cm component of Fig.~\ref{fig:hi}[B] is clearly detected in 
both observing runs, with the correct Doppler shift due to the Earth's motion; 
there is no doubt about its reality. While the weaker 21cm component (at 
$\sim +45$~\kms) was only detected in the higher sensitivity spectrum of August~2005, 
the data showed no evidence for RFI and the feature is detected at $> 5\sigma$ 
level. Moreover, as we will discuss later, both the redshift and the velocity width 
of this feature are in excellent agreement with those of a component 
detected in the FeII~ $\lambda$2374 transition. We conclude that the second 
21cm component is also likely to be real. 

The 21cm absorption towards \pks~extends, between nulls, over $\sim 60$~\kms,
with an equivalent width of $\sim (0.216 \pm 0.027)$~\kms. The solid line 
in Fig.~\ref{fig:hi}[B] shows a two-component fit to the spectrum, with 
the components modelled as gaussians. The peak redshifts of the components 
are $z_1 = 2.347477 (12)$ and $z_2 = 2.347869 (20)$ and their peak optical depths 
(the ratio of line depth to continuum flux density, using the low optical depth limit; 
\pks~is unresolved by the GBT beam) are $\tau_1 = (6.2 \pm 0.8) \times 10^{-3}$ 
and $\tau_2 = (3.3 \pm 0.7) \times 10^{-3}$ (including errors in 
the source flux density). They have fairly large velocity spreads 
(FWHMs of $\sim (23.3 \pm 2.7)$~\kms~and $\sim (18.2 \pm 4.5)$~\kms, 
respectively), suggesting either a blend of many cold narrow components or that 
bulk/turbulent motions are important. Alternatively, if the absorption arises in a 
warm phase, with the velocity spread primarily due to thermal motions, the implied 
kinetic temperatures of the two components are $\lesssim (11865 \pm 2750)$~K 
and $\lesssim (7240 \pm 3580)$~K. 

Fig.~\ref{fig:co} shows the final ATCA redshifted CO~3--2 spectrum towards \pks, 
with flux density, in mJy, plotted against barycentric velocity, in \kms, relative to 
$z = 2.34736$; the dashed line indicates the peak of the 21cm absorption. 
The spectrum has been smoothed to (and resampled at) a velocity resolution 
of $\sim 7.25$~\kms~and shows no evidence for absorption, with an RMS noise 
of $\sim 8.6$~mJy per 7.25~\kms. The $3\sigma$ upper limit on the optical 
depth is $\tau < 0.058$, per $\sim 10.2$~\kms. Curiously, however, a weak 
emission feature can be seen in the spectrum at the redshifted CO~3--2 
frequency. While this has only $\sim 3\sigma$ significance in a single 
channel (after smoothing by 3~channels and resampling),
it is somewhat tantalizing, especially as it is present (albeit at even weaker 
levels) in the two independent polarizations and occurs at the 21cm redshift. 
We discuss the implications of a positive detection in the next section.

\section{Discussion}
\label{sec:discuss}

\begin{figure*}
\centering
\epsfig{file=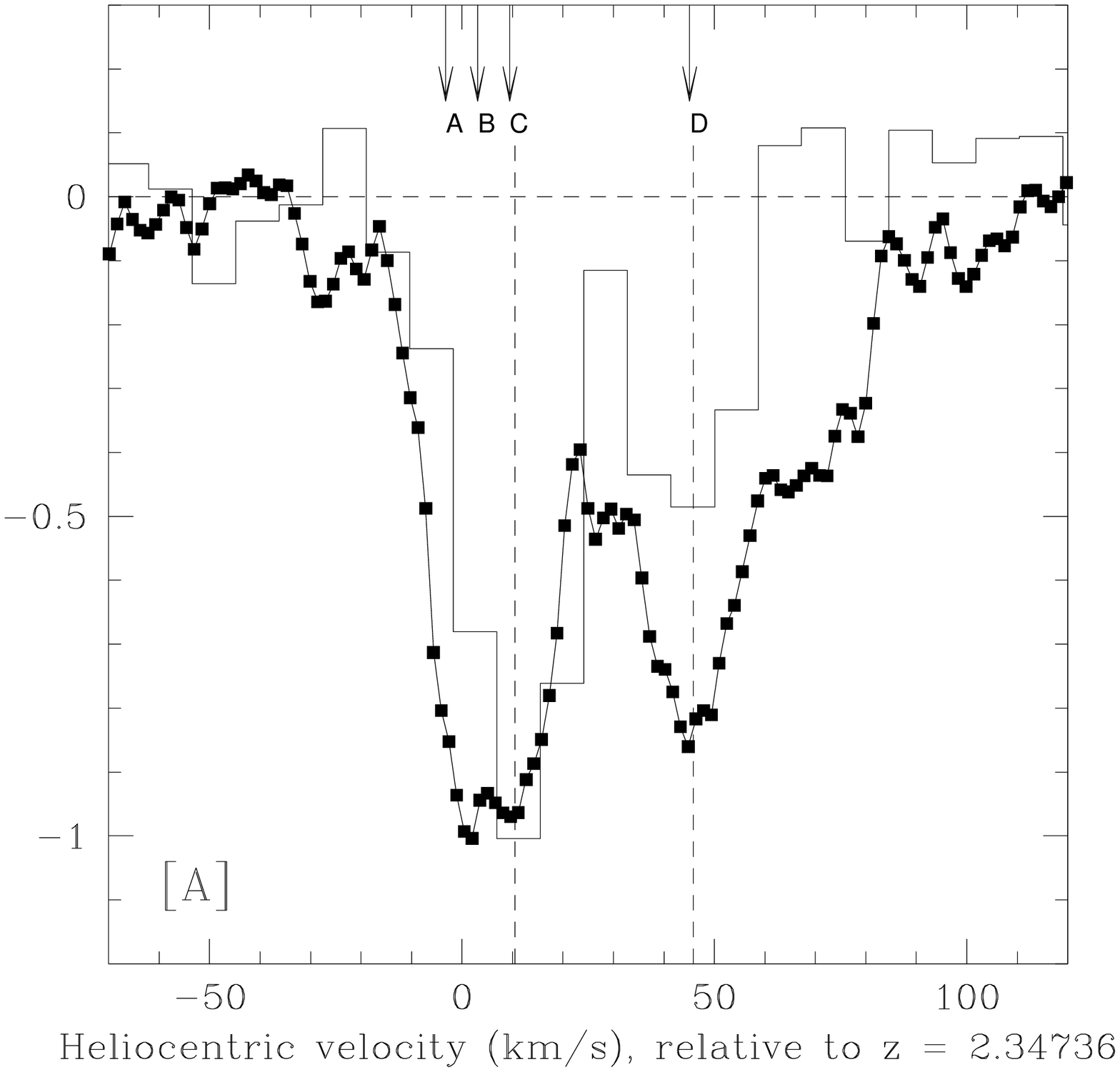,height=3.3truein,width=3.3truein}
\epsfig{file=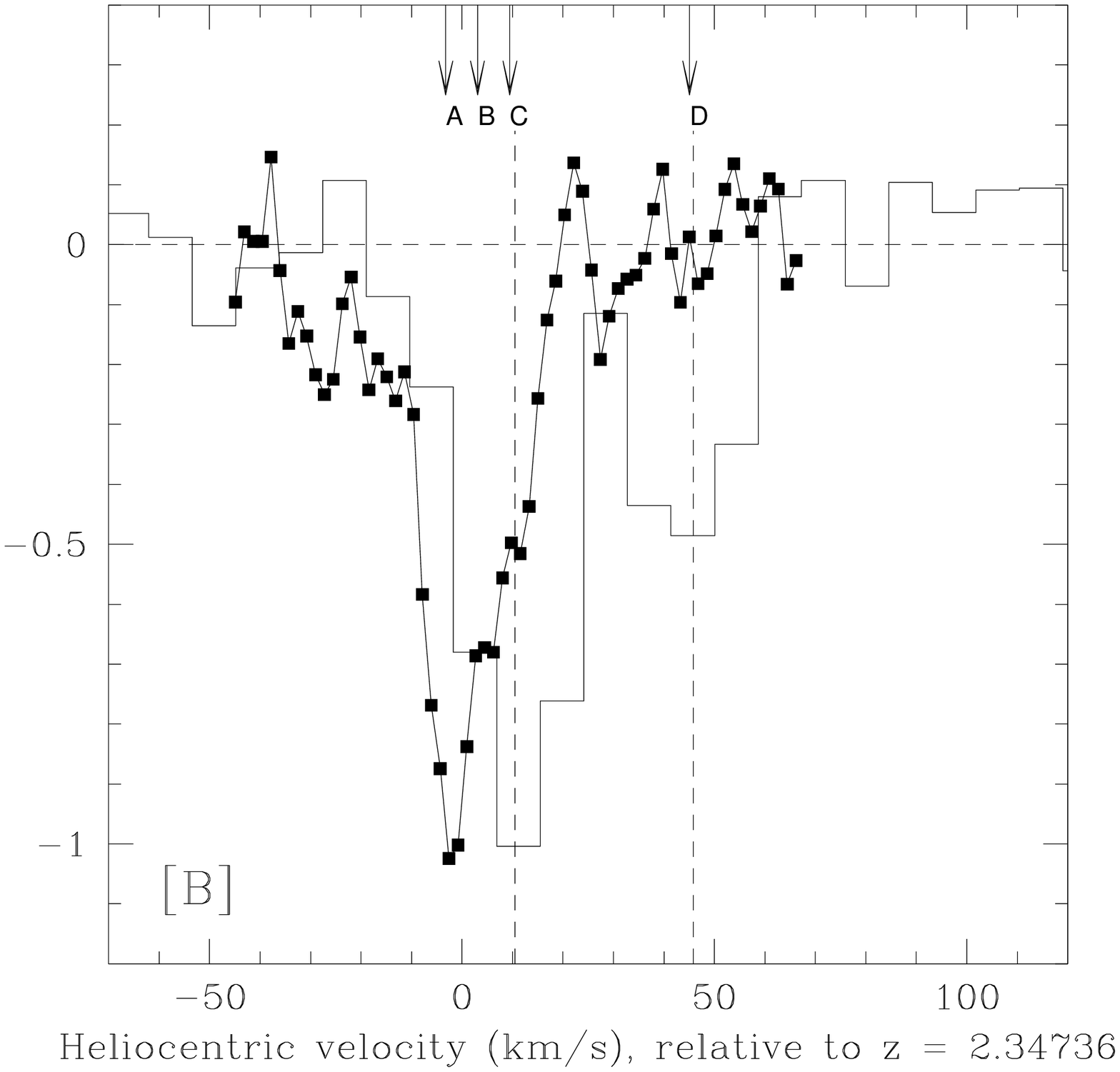,height=3.3truein,width=3.3truein}
\vskip -0.1in
\caption{A comparison between the GBT HI~21cm profile of Fig.~\ref{fig:hi}[B] 
(histogram) and the [A]~FeII~$\lambda$2374 and [B]~ZnII~$\lambda$2062 profiles 
(solid points), with all spectra in arbitrary units. The redshifts of the two 
21cm components are shown by the two dashed vertical lines while the 
arrows at the top of each panel (marked A, B, C and D) indicate the 4 FeII components 
of Table~1. See text for discussion.}
\label{fig:hi-metal}
\end{figure*}

For optically thin 21cm absorption, the HI column density $\NHI$ (in \cm), 
21cm optical depth $\tau_{21}$ and spin temperature $\ts$ (in K) 
are related by the equation 
\begin{equation}
\NHI = 1.823 \times 10^{18} [\ts / f] \int \tau_{21} {\rm dV} \;\; ,
\end{equation}
where the profile integral is over velocity (in \kms) and the covering factor 
$f$ gives the fraction of the radio source that is obscured by the foreground 
cloud. The detection of 21cm absorption in a DLA, where the HI column density is 
known from the Lyman-$\alpha$ line, thus allows an estimate of the spin 
temperature of the absorbing gas (assuming that the HI column density 
measured towards the optical QSO is the same as that towards the radio source). 
As has often been emphasized (e.g. \citealt{carilli96,chengalur00}), the spin temperature 
thus obtained is the column-density-weighted harmonic mean of 
the spin temperatures of different HI phases along the line of sight. 
In the case of the $z \sim 2.347$ DLA, the 
HI column density is $\NHI = (6 \pm 1.5) \times 10^{20}$~\cm~\citep{ellison01}. 
The measured 21cm equivalent width of $\int \tau_{21} d{\rm V} = 
(0.216 \pm 0.027)$~\kms~then yields $\ts = [(1527 \pm 428) \times f]$~K, 
where the error is dominated by the uncertainty in $\NHI$. 

In order to estimate the covering factor, it is important to first determine 
whether both 21cm absorption components of Fig.~\ref{fig:hi}[B] 
arise against the quasar core. Figs.~\ref{fig:hi-metal}[A] and [B] show 
a comparison between the HI~21cm profile (histogram) and the 
FeII~$\lambda$2374 and ZnII~$\lambda$2062 profiles of \citet{akerman05}. 
The FeII and 21cm profiles are similar, with FeII absorption seen at 
the velocities of both 21cm components; this suggests that the latter
indeed arise from the line of sight towards the optical nucleus.

The covering factor $f$ can thus be assumed to be the fraction of source flux density 
contained in the core and is best estimated from VLBI observations of the 
radio source at the redshifted 21cm line frequency; unfortunately, no 
such observations of \pks~exist in the literature. The lowest frequency 
with published VLBI observations is 2.3~GHz, where most of the source flux 
density stems from two compact components, separated by $35$~mas \citep{preston89}. 
Of these, the northwest component is more compact than the southeast one 
($\sim 5$~mas v/s $\sim 18$~mas) and is likely to be the nuclear core, especially 
since it is also very compact in higher resolution ($\lesssim 2 $~mas) 4.8~GHz VLBI and 
VSOP images \citep{shen98,tingay02}. The core has a flux density of $\sim 2.5$~Jy 
in the 2.3~GHz VLBI image, $\sim 58$\% of the total 2.3~GHz flux density 
\citep{preston89}. Unfortunately, the source is highly variable and it is thus 
not possible to determine its spectral index or its core flux density at 424.3~MHz
from the 2.3 and 4.8~GHz VLBI data, as these observations were carried out 
at very different epochs, many years before our observations.
For now, we are forced to estimate the covering factor solely from the 2.3~GHz VLBI 
information. Assuming that only the core is completely covered gives $f = 0.58$ from 
the 2.3~GHz data, and, hence, a high spin temperature, $\ts = (886 \pm 248)$~K. 
Of course, it is quite possible that both 2.3~GHz source components 
are covered by the DLA, since their separation of $35$~mas corresponds to a 
linear scale of only $\sim 290$~pc at $z = 2.347$ (using an LCDM cosmology, with
$\Omega_m = 0.3$, $\Omega_\Lambda = 0.7$ and $H_0 = 70$~km/s~Mpc$^{-1}$). 
In such a situation, with $f\sim 1$, the spin temperature would be even higher, 
$\ts \sim (1527 \pm 428)$~K. Conversely, it is possible that the spin 
temperature is significantly lower than this, if the core has an inverted 
spectrum and is the only source component covered by the DLA. We note that the 
other confirmed $z > 1$ HI~21cm absorbers, in the DLAs towards QSO~1331+170, 
PKS~1157+014 and PKS~0458$-$020, also have relatively high spin temperatures, 
$\ts \gtrsim 500$~K (e.g. \citealt{kanekar03}).

\begin{table}
\label{tab:metals}
\begin{center}
\begin{tabular}{@{}|c|c|c|c|c|c|c|}
\hline
      & Velocity & Redshift & $b$    &N$_{\rm  FeII}$  & N$_{\rm ZnII}$ &  [Zn/Fe] \\
      & \kms     &    z     & km/s &\cm	& \cm & \\
\hline
&&&&&& \\
A	& $-3.2$ & 2.347324 & 1.0 & 14.58 & 12.07 & 0.33  	\\
B	& $+3.1$ & 2.347395 & 9.0 & 14.23 & 12.54 & 1.15 	\\
C	& $+9.5$ & 2.347466 & 9.3 & 14.34 & 11.83 & 0.33 	\\
D	& $+45.1$& 2.347864 & 11.3 & 14.28 & $< 11.7$ & $< 0.26$ \\
&&&&&& \\
\hline
\end{tabular}
\caption{The FeII and ZnII column densities and [Zn/Fe] for the four 
strongest FeII components (in logarithmic units), with the [Zn/Fe] values 
relative to the solar abundances of \citet{lodders03}. Components
C and D have velocities in good agreement with those of the 21cm components.}
\end{center}
\vskip -0.15in
\end{table}

The average metallicity of the $z \sim 2.347$ DLA is quite high for these redshifts, 
[Zn/H]~$=-0.68 \pm 0.15$~\citep{akerman05}. Of course, the lack of velocity information 
in the Lyman-$\alpha$ line precludes measurements of the HI column densities (and hence, 
the metallicities) of individual absorption components. We have used VPFIT to fit 
the ZnII and FeII transitions detected in the VLT-UVES spectrum of \citet{akerman05}; 
this has a resolution of 
$\sim 7$~\kms~and S/N~$\sim 25$ near the ZnII and FeII wavelengths. Table~1
lists the velocities (relative to $z = 2.34736$), redshifts, $b$-parameters, FeII and ZnII
column densities and the dust depletion (as measured by [Zn/Fe]) of the four 
strongest components in the DLA, obtained from the fits. The first three 
components (A -- C) in the table contribute to the strongest optical absorption (peaking 
at $v \sim 0$~\kms~in Fig.~\ref{fig:hi-metal}, $\sim 11$~\kms~ blueward of the 
stronger 21cm absorption), 
while the fourth is associated with the weaker 21cm absorption, at $v \sim 45$~\kms. 
The weaker 21cm component is not detected in the ZnII lines; $3\sigma$ upper limits 
on its ZnII column density and dust depletion are listed in Table~1. Although the optical 
absorption has been decomposed into sub-components during the fitting
procedure, the modest S/N of the UVES spectrum and the complex kinematics
of the metal lines introduce a considerable amount of degeneracy.
Errors on the column densities and redshifts of individual components
are therefore sizable.  However, the total column densities are well
constrained, with our values in excellent agreement with those of
\citet{akerman05}, despite the very different fits.
We hence combine the three optical components at $v \sim 0$~\kms~to 
obtain [Zn/Fe]$= (0.62 \pm 0.14)$, indicating a fairly high dust depletion 
compared to other DLAs at this redshift. Conversely, the 
component at $v \sim 45$ km/s has a much lower implied dust-to-gas ratio, 
[Zn/Fe]$< 0.26$. Fig.~8[A] of \citet{wolfe05} shows that all DLAs 
with [Zn/Fe]$ < 0.4$ have [Zn/H]$ \lesssim -1$.  Although this result is 
obtained from [Zn/H] and [Zn/Fe] values averaged over the entire profile 
(i.e. not from individual components), it suggests that the $v \sim 45$~\kms~gas 
has a low metallicity. The $v \sim 0$~\kms~gas should then have a 
fairly high metallicity, to account for the high average metallicity of 
the absorber. The DLA thus appears to contain some gas that is rich in 
dust and metals and with a relatively large 21cm optical depth, as well 
as a second component that is poorer in heavier elements and with weaker 
21cm absorption. The very different depletions seen in the two components 
imply that the absorber has a rather non-uniform interstellar medium, as 
recently seen in other high $z$ DLAs \citep{dessauges05}. In fact, the 
FeII~$\lambda$2374 and ZnII~$\lambda$2062 lines at $v \sim 0$~\kms~have 
quite different shapes, suggesting that even the sub-components within 
this complex have significantly different physical conditions.

Comparisons between 21cm and optical redshifts in a statistically large absorber
sample can be used to probe the evolution of fundamental constants \citep{wolfe76}.
The most sensitive result is that of \citet{tzanavaris05}, who compare the 
redshifts of strongest 21cm and metal-line absorption in 8 DLAs to 
constrain changes in the quantity $x \equiv g_p \mu \alpha^2 $, where $\alpha$ is the 
fine structure constant, $g_p$, the proton gyro-magnetic ratio and 
$\mu \equiv m_e/m_p$, the electron-proton mass ratio. However, it is by no
means essential that the strongest 21cm and metal absorption both arise from 
the same absorption component; for example, \citet{chengalur00} noted that
the deepest UV absorption in the $z \sim 2.04$ DLA towards PKS~0458$-$020 is 
offset from the strongest 21cm component but in good alignment with the secondary 
21cm component. Similarly, both panels 
of Fig.~\ref{fig:hi-metal} show that the strongest optical absorption 
in \pks~(at $z \sim 2.347360$) is offset from the strongest 21cm absorption 
by $\sim 11$~\kms. In the approach of \citet{tzanavaris05}, these offsets would 
imply either evolution in the fundamental constants or an intrinsic velocity 
offset between the 21cm and metal lines. However, it can be seen from 
Fig.~\ref{fig:hi-metal}[A] 
and Table~1 that there are weaker FeII components (C and D) in \pks~much closer 
to the redshifts of both 21cm components. While component~C (at $v = +9.5$~\kms) 
is severely blended with the complex at $v \sim 0$~\kms, making it difficult to 
determine an accurate redshift from the present spectrum, the fit to component~D 
at $v \sim +45$~\kms~is quite stable, yielding the redshift $z_{1} = 2.347864 (5)$ 
(note that the RMS-error of the wavelength solution is $\sim 250$~m~s$^{-1}$, 
a factor of two smaller than the error from the fit).
Comparing this to the redshift of the nearest (weaker) 21cm component, 
$z_{2} = 2.347869 (20)$ (from the 2-gaussian fit), we obtain $[\Delta x/x] = 
\Delta z/(1 + {\bar z}) = (-0.15 \pm 0.62) \times 10^{-5}$, where 
$\Delta z = z_{1} - z_{2}$ and ${\bar z}$ 
is the mean of the redshifts; this is consistent with the null result of no evolution.
Note that this error does not include possible systematic velocity offsets between 
the 21cm and UV redshifts, which can only be addressed with statistically large 
samples. However, it is clear that the results obtained from such comparisons 
between different species (and using lines at very different wavelengths) 
critically depend on the precise details of the comparison. The assumption that 
the strongest absorption in the two species arises in the same component introduces 
an extra source of error, that could well dominate the $\sim 6$~\kms~ 
``intrinsic'' offset between 21cm and UV redshifts obtained by \citet{tzanavaris05},
leading to a bias in favour of a detection (especially in small samples). It may 
thus be better to compare the redshifts of the {\it nearest} simple (i.e. unblended) 
metal and 21cm components, after using high resolution spectroscopy to decompose the 
profiles into their components, as has been done here. While this might 
have the opposite bias, towards a non-detection of evolution, it could be 
considered a test of the null hypothesis. In any case, care must 
be taken in the interpretation of the results from such comparisons.

The ATCA non-detection of CO~3--2 absorption in Fig.~\ref{fig:co} places limits
on the CO column density in the $z \sim 2.347$ DLA. Assuming a velocity width 
of 10~\kms~gives the $3\sigma$ limit N$_{\rm CO} < 3.8 \times 10^{15} \times (T_x/10)$~\cm, 
similar to earlier limits in DLAs (e.g. \citealt{curran04b}).  A conversion factor 
of $10^5$ from N$_{\rm CO}$ to N$_{\rm H_2}$ \citep{liszt00} gives N$_{\rm H_2} < 3.8
\times 10^{20} \times (T_x/10)$~\cm, i.e. a molecular fraction $f < 0.6$.

The possible CO~3--2 emission feature seen in Fig.~\ref{fig:co}, if real, corresponds 
to a molecular mass of M$_{\rm H_2} \sim 4.2 \times 10^{10} \:\: {\rm M}_\odot$, 
an exceedingly large value (using the Galactic conversion factor of 
$4.6 \:\: {\rm M}_\odot {\rm (K \:\: km/s \:\: pc^2)^{-1}}$ from the CO~line 
luminosity $L'_{\rm CO}$ to mass; \citealt{solomon97}). CO emission has never 
before been detected from a DLA. We note that the velocity width of the feature is 
quite narrow, $\sim 30$~\kms;  if real, this would be surprising for 
such a large molecular mass and would require the galaxy to be quite 
close to face-on to the line of sight. 

In summary, we have used the GBT to detect HI~21cm absorption in the 
$z \sim 2.347$ DLA towards \pks. We obtain a high spin 
temperature $\ts = (886 \pm 248) \times (f/0.58)$~K, 
estimating the covering factor $f$ from 2.3~GHz VLBI observations. 

\vskip 0.2in
\noindent {\large \bf Acknowledgements}\\
We thank Bob Garwood for much help with the AIPS++ data analysis and 
Carl Bignell and Toney Minter for help with the GBT observations. 
%Part of this work was completed while NK was a Visiting Scientist at ESO, 
%Santiago and a visitor at the ATNF, Sydney; he thanks ESO and the ATNF 
%for hospitality. 
Basic research in radio astronomy at the Naval Research Laboratory is
funded by the Office of Naval Research. The NRAO is operated by 
Associated Universities, Inc., under cooperative agreement with 
the National Science Foundation. The ATCA is part of the Australia Telescope, 
funded by the Commonwealth of Australia for operation as a 
National Facility managed by CSIRO.

\bibliographystyle{mn2e}
\bibliography{ms}

\end{document}